\documentstyle[aps,prl,epsf]{revtex} 

\begin{document}
\twocolumn[{  
\draft        
\title{\bf 
Anomalous Transport in Conical Granular Piles
}
\author{
Peter Ahlgren$^1$\cite{underg},
Mikkel Avlund$^1$\cite{underg},
Ib Klewe$^1$\cite{underg},
Jonas Nyvold Pedersen$^1$\cite{underg},
and
\'Alvaro Corral$^{1,2}$
\cite{email} 
}
\address{
$^1$
The Niels Bohr Institute, University of Copenhagen,       
Blegdamsvej 17,  DK-2100  Copenhagen \O,   Denmark \\
$^2$
Departament de F\'\i sica, 
Universitat Aut\`onoma de Barcelona,
Edifici Cc, E-08193 Bellaterra, Barcelona, Spain 
}
\date{Accepted in PRE 15 July 2002; published ???}
\maketitle 
\widetext  
\begin{abstract}
\leftskip 54.8 pt 
\rightskip 54.8 pt 
Experiments on $2+1$-dimensional piles of elongated particles are performed.
Comparison with previous experiments in $1+1$ dimensions 
shows that the addition of one extra dimension to the dynamics 
changes completely the avalanche properties,
appearing a characteristic avalanche size.
Nevertheless, 
the time single grains need to cross the whole pile
varies smoothly between 
several orders of magnitude, from a few seconds to more than 100 hours.
This behavior is described by a power-law distribution,
signaling the existence of scale invariance in the transport process.
\end{abstract}
\leftskip 54.8 pt 

\pacs{
PACS numbers:  
45.70.Ht,   
64.60.Ht,   
05.60.Cd,
05.65.+b    
}
}]   

\narrowtext
\setcounter{page}{1} 
\markright{\bf Anomalous Transport in Conical Granular Piles.
Accepted in PRE} 
\thispagestyle{myheadings} 
\pagestyle{myheadings} 



\section{ Introduction.}
Transport phenomena are found at the core of almost every
discipline in many-body physics;
for example 
in nuclear physics, 
with the motion of neutrons through 
a fissionable material;
in electronics, 
with the transport of holes and electrons in semiconductors;
in astrophysics, 
with the diffusion of light through stellar atmospheres;
and also 
in plasma physics, biological physics, etc. \cite{Duderstadt}.
The study of granular matter 
(like sand, powders, and all that)
has only very recently become
a goal for
the physics community (surprisingly, see Ref. \cite{Behringer})
and therefore
the understanding of their transport properties
will be required in order to explain the (sometimes-strange) properties
of these materials \cite{reviews,Duran}.

Seminal experimental work on transport of individual grains
through a granular pile was performed by Christensen {\it et al.} 
\cite{Christensen}:
a $1+1$-dimensional pile was built in the narrow gap
between two parallel plates 
using elongated rice grains.
The time taken by tracer grains to cross the system
was measured
when the pile was driven out of equilibrium 
by a constant addition of grains.
The probability density of these {\it transit times} turned out to be,
for long times,
a decreasing power law with an exponent $\alpha \approx 2.4$,
signaling the existence of an anomalous (non-Gaussian) behavior.
By means of some theoretical models
the transport process was later microscopically understood 
as a composition of L\'evy flights (in the jumps of the
grains during an avalanche) and also power-law distributed
trapping times \cite{Boguna}. 

On the other hand, the same rice pile was previously shown 
to display self-organized criticality (SOC) 
due to the dissipation introduced in the motion by the elongated
shape of the grains, which removes inertial effects \cite{Frette}.
This SOC behavior simply means that 
the motion of the grains takes place in terms of avalanches, 
and the distribution of sizes of these avalanches is scale invariant, 
that is, another power law,
achieved without any parameter fine tuning \cite{Bak}.
Interestingly, SOC does not appear when the grains are somewhat round,
and therefore more free to roll and accumulate kinetic energy \cite{Frette}.

Additionally, other remarkable properties of this rice-pile system include 
diverging profile fluctuations with system size \cite{Malthe},
and
probably a multifractal spectrum of the transit-time time series \cite{Pastor}.
More generically, a pile configuration
allows to study the transition between different metastable
configurations of a granular system,
which occurs when a solid phase becomes unstable
and flows like a liquid until it becomes "frozen"
again.

Another class of experiments, performed on
sandpiles (or with round particles, in general)
in $2+1$ dimensions show that
in these systems a crossover from small
to large avalanches occurs when the system size is increased 
\cite{Held,Rosendahl,Feder};
that is, small piles tend to show avalanches of
different sizes, although not very large, 
but for a big enough system  
very large avalanches appear in addition,
a behavior that is not compatible with SOC
\cite{Corral}.
This is in agreement with a different type of experiments
performed on rotating drums \cite{Jaeger,Nagel}.

The extension of rice-pile experiments to $2+1$ dimensions,
which is the subject of our research,
allows a comparison with the results for sandpiles
and raises
a series of interesting questions:
will the dissipation of the elongated grain motion 
be enough to suppress inertia and give SOC?
If not, which is then the behavior?
But even more important, 
our experiments allow 
to explore the transport of individual
grains in granular piles,
and in general
to go deeply
into the connection between anomalous transport and SOC,
a connection that has not been explored in detail.
We find that 
the behavior 
of the avalanches in a conical rice pile (with elongated grains)
is very similar to what has been observed in conical sandpiles
\cite{Held},
with somewhat small avalanches for small piles
and large relaxation oscillations in larger systems.
However, the transport properties give totally new information,
with a power-law transit-time distribution even in the later case
of almost periodic occurrences.

\section{ Experimental setup and procedure.}
The rice pile was built over a wooden (circular) disk
supported by a platform which allowed the grains falling out
in any direction when they reached the disk perimeter.
Three different disk diameters were employed, 
$L=10, 15$, and $20$ cm,
in order to study the effect of the system size.
In terms of the mean grain length (6.6 mm, see below)
this quantity turns out to be $L=15$, 22.5, and 30.

The driving of the pile was carried out by means of a single-seed machine
connected to a DC motor through a gearbox.
The driving rate was adjusted with a voltage transformer
to a value of 43 grains per minute ($2 \pm 0.4$ grains every 2.8 seconds);
this rate is small enough to be considered as a zero driving rate
for the system sizes studied (see Ref. \cite{Corral99}) 
but also allows to conclude the experiment in a finite time.
After being expelled from the machine 
the grains were directed by a cardboard pipe towards a funnel
where at the exit a paper tongue braked the grains and dropped
them over the center of the disk with few kinetic energy.
Once the pile was built, the distance from the
exit of the funnel to the top of the pile 
was just a few centimeters. 

In addition to the described driving procedure,
when the pile reached a (statistical) stationary state
--where the injected grains were balanced out on average
by the coming out grains--
tracer grains were added to the top of the pile by hand
at a rate of 1 grain every 30 seconds
for the two smallest piles and 1 grain every minute
for the largest one \cite{driving}.
These tracer grains were used to compute the transit time,
which is the time a grain remains in the pile,
since it is added until it comes out.
Tracers were marked by a number written on the surface,
and were also colored in order to be visually distinguished
from the rest of the grains.
The addition of each tracer was done at a fixed input time
$T_{in}$
(determined by the tracer driving rate),
being only necessary then to record the output time 
$T_{out}$ 
to get the transit time as 
$T=T_{out}-T_{in}$.
The total number of injected tracers was 
about 500 for $L=15$ and $L=22.5$ and 700 for $L=30$.
After the end of the tracer injection,
the experiment and therefore the normal driving of the pile
should continue until all tracers come out;
however, for practical reasons this could not be achieved
and the experiment was stopped when a few tracers were
still inside the pile, after more than 100 hours.
The elimination of these long-living tracers
affects the direct calculation of the mean transit time,
but not the whole probability distribution, which is a much more
interesting quantity,
as we will see.

In practise, the grains coming out from the pile 
were collected by a piece of cardboard
and directed at a glass, where the colored tracers were
easily recognized by the naked eye.
It was then when the output time was recorded,
and the glass replaced by another one.
Moreover, note that the grains collected in the glass after an avalanche
can give a measure of the avalanche size (of course for those avalanches
that reach the boundaries).
During the addition of the tracers three persons were needed
to perform the described tasks, but as long as the number of tracers
inside the pile decreases these demands diminish too.

The rice used in all the experiments was "L\o se parboiled ris"
from F\o tex (Denmark). The shape of these grains is elongated,
similar to the one of Refs. \cite{Frette,Christensen},
with a length of $6.6 \pm 0.9$ mm and a width of $1.6 \pm 0.6$ mm,
which gives a length-width ratio of about 4.
The mean weight of the grains is $16.9$ mg.

\section{ Experimental results.}
%
The first observation provided by the experiment is a qualitatively one:
as in sandpiles, 
the size of the system influences dramatically the behavior.
Small piles ($L=15$ and $22.5$) do not change very much their profile
during time evolution once the pile is built 
(the total heigh varies in about 1 cm)
and do not display big avalanches either. 
In contrast, the big pile ($L=30$) shows enormous, 
catastrophic avalanches,
which eject a significative percentage of the grains out of the pile
and therefore change totally the profile.
In the time interval in between these enormous avalanches 
there are also small avalanches.
Actually, this is a similar behavior to what was observed 
in sandpiles (with round particles) 
\cite{Held}.
Therefore, the elongation of the rice grains 
(which was identified as the responsible for SOC in $1+1$-d \cite{Frette})
does not make dissipate
enough energy in three dimensions to give SOC.
The addition of one extra degree of freedom to the motion of the grains
modifies totally the behavior, 
in comparison.

To explain the transition in sandpiles when the size of the system
is increased, a simple argument was proposed  
based on the fact that for small piles there is "no room"
for the two angles involved in the dynamics
(one for the starting of the avalanche and another one for the end)
\cite{Nagel}.
We find that the situation is not so easy,
due to the somewhat rough surface of the pile;
indeed, the elongation of the grains gives rise to local bumps of packed grains
which provoke fluctuations of the total height of the pile.
Moreover, in the regime with large avalanches the measurement 
of the angle of repose --the angle after an avalanche--
makes little sense, since then a large part of the pile has collapsed
(like the crater of a volcano) and the conical shape is lost.
Rotating-drum experiments could perhaps give more information
on this point.

The total mass of dropped grains
and the occurrence time 
of these catastrophic avalanches 
(in the big pile) was measured.
In all cases the number of dropping grains was in between about 3000 and 7000,
with a mean value of 4600 and a standard deviation of 1100.
For comparison, the largest avalanches in the small piles dropped only about
20 or 30 grains,
roughly the same value as the largest small avalanches in the big pile.
The mean time between big avalanches was 7700 grains (3 hours)
with a deviation of 2200.
The difference with respect the mean number of dropping grains in big avalanches (4600)
corresponds of course to grains coming out in small avalanches.

%
\begin{figure}
\epsfxsize=2.9truein 
\hskip 0.15truein\epsffile{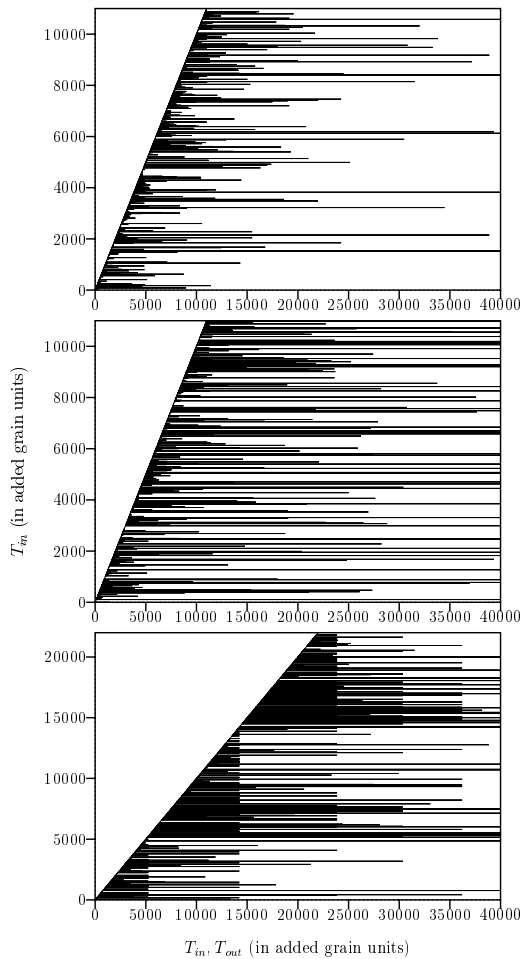} 
\caption{ 
Sequence of input and output times for different tracers and 
the three system sizes studied,
$L=15$, $22.5$, and $30$, from top to bottom.
Different tracers are represented by different lines,
the starting point is the input time,
the end is the output time, and the transit time
is represented by the length.
Zero time corresponds to the starting of the tracer injection.
Time units  
are set by the driving rate,
the window shown in the x-axis corresponding to $15.5$ hours.
Notice that for the big pile 1 out of 44 grains is a tracer,
whereas for the other two the proportion of tracers is double
(2 out of 45); 
therefore the plots understimate the difference between both behaviors.
\label{sew}
}
\end{figure}
%

Let us address now the problem of the transport inside the system.
Part of the data for the input times and their corresponding output times
for the three experiments are shown in Fig. \ref{sew},
where transit times are given by the length of the horizontal lines.
First, it is clear the difference between small and big systems,
in terms of the number of coincidences in the output times
(related to the avalanche size).
Moreover, although comparison with Christensen {\it et al.}'s
results \cite{Christensen} is not direct (tracer density is lower there),
it is possible to see that their Fig. 2 settles in between
our two behaviors.
Second, one can observe the great variability of the transit times,
broadly distributed from few seconds to more than 100 hours
(this last value, not shown in the plot,
is in fact a lower bound for the maximum transit 
time, determined by the duration of the experiment).

%
\begin{figure}
\epsfxsize=2.9truein 
\hskip 0.15truein\epsffile{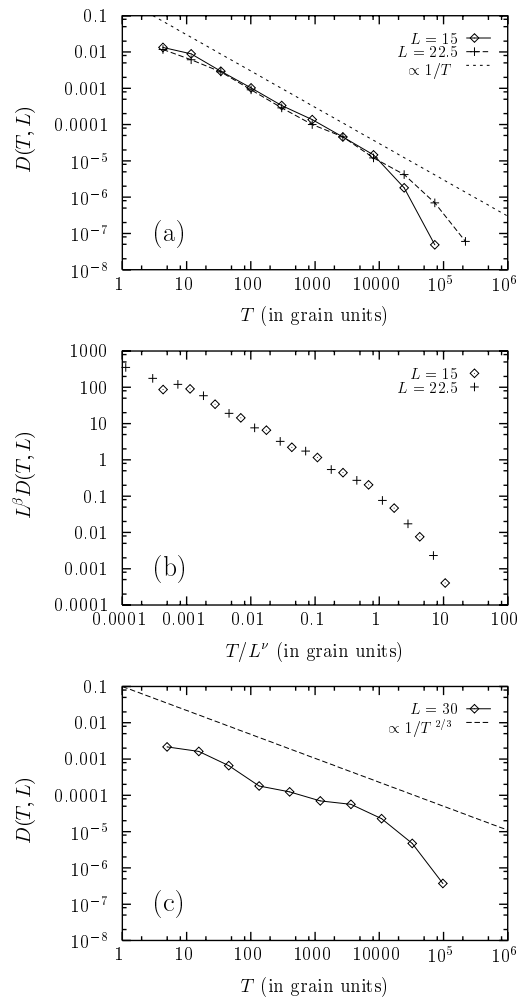} 
\caption{ 
Probability densities corresponding to the transit times 
for the system sizes studied,
calculated over exponentially increasing intervals of size
$b^n$.
As usual, time is measured in terms of added grains,
and $L$ in terms of grain length.
Times smaller than 3 seconds are not included.
(a) The two smallest piles, using $b=2.5$.
The straight line is a decreasing power law with exponent 1.
(b) Finite size scaling of the previous distributions,
with $b=2$ now.
(c) The big pile, with $b=2.5$ again.
A power law with exponent $2/3$ is also shown.
\label{Dtrans}
}
\end{figure}
%

Figure \ref{Dtrans} displays the probability densities of the transit times measured
for each pile.
It is apparent once more the difference between the two smallest piles and the
big one.
In the first case (Fig. \ref{Dtrans}(a)) 
we obtain a power-law distribution of transit times
with an exponent close to one, i.e.,
$
D(T,L) \propto 1/T^{\alpha}, \mbox{ with } \alpha \approx 1,
$
ranging from tens of seconds to tens of hours, over more than 3 decades.
This is a signature of anomalous diffusion and dispersive transport,
and it has been widely studied in amorphous semiconductors, glasses,
and many other disordered materials \cite{Scher}.

This self-similar behavior is limited by a sharper cutoff for long times,
which increases with system size.
Finite size scaling, shown in Fig. \ref{Dtrans}(b) ratifies this,
giving the relation
$
D(T,L) = {L^{-\beta}} f({T}/{L^{\nu}})
$
between the distributions for different sizes of the system.
The data collapse gives for the exponents 
$\beta \approx \nu \approx 3.3 \pm 0.5$,
in agreement with 
a well known scaling relation (obtained from the normalization condition),
$\beta=\nu$ if $\alpha \le 1$.
Also, from these relations 
it is easy to obtain that the mean transit time scales as
$ \langle T(L) \rangle \propto L^{\nu} $,
which is a very fast increase.
This procedure is more practical than direct calculation 
of $ \langle T \rangle $,
since it can be used
when data for some extreme events are unavailable,
as we have for very deeply buried tracers.
Notice also that a generalization of the argument of 
Ref. \cite{Christensen} would give an active zone increasing as
$L^{\nu-2}$, so, even in the regime of small avalanches
there is an active zone that increases with system size.

However, further increments on the size $L$ break completely this behavior
and the previous scaling relations are no longer valid.
Indeed, for the big pile we enter into the regime of large relaxation
avalanches and the transport properties consequently change.
As we see in Fig. \ref{Dtrans}(c) the transit time distribution becomes 
even broader, as it decays more slowly.
A decreasing power law with exponent $2/3$ is shown for comparison,
although the statistics is not so good as for the small piles,
due to the extremely slow decay.
We find particularly relevant that despite of the almost periodic
ticking of the pile, there is no characteristic scale for the 
transit times.

\section{ Conclusions and perspectives.}
%
As we have seen,
three-dimensional piles of elongated particles behave essentially as piles of
round particles in 3d,
rather than as $1+1$-dimensional rice piles.
Therefore, the transport properties we have found should apply
to any sandpile in 3d \cite{Held,Rosendahl}.
It is an open question if with an even more dissipative particle shape,
piles in 3d would give scale invariance in the distribution of avalanches.
In fact we believe that the use of particles with several "arms",
like crosses or 3d crosses, or some kind of sticky particles,
would brake the tendency to roll and the results
--if this kind of particles could be experimentally handled with success--
would be closer to SOC.

However, although we do not see indication of SOC in the distribution of avalanches,
we find that the transit time of the grains is power-law distributed,
signaling the existence of a scale-free behavior,
as in the $1+1$ dimensional SOC rice pile of Refs. \cite{Frette,Christensen}.
Nevertheless it is worth mentioning that the type of distributions that we find
are much broader (in the sense that the exponent $\alpha$ is much smaller)
than those of the $1+1$-d case.

Big piles, that is, those in the relaxation-oscillation regime,
demmand a more intensive study.
Notice that we have studied the smallest of these big piles,
just above the transition, but we have found enormous transit times
which provoke serious inconveniences in the experimental realization.
Indeed, the fact that the transit time distribution decays more
slowly than for the small piles, gives a larger number of tracers in the
tail of the distribution, which corresponds to long-living particles.
This effect increases dramatically with system size;
therefore, much better statistics is necessary for the elucidation 
of the characteristics of the transport process.
We are talking about of tens of thousands of tracers,
which requires a different procedure than our hand addition and visual
control, so man-time consuming.
In consequence, automatization of the tracer recognition process would be
highly desiderable.

Finally, none of the sandpile models we know so far
is able to reproduce these results;
accordingly, it would be of the maximum interest to develop models
that allow the necessary interplay between theory and experiment
to overcome our present knowledge of these media.

\section{ Acknowledgements.}

The experiment was designed and performed by the first four authors
as first-year undergraduate students 
of the University of Copenhagen,
for a project for the course Fysik I
with the assistance of the last author.
The realization was possible thanks to the enthusiastic support
of Anders Holm.
A.C. is indebted to the Cooperative Phenomena Group of the 
University of Oslo, 
where he was introduced into the wonderful world of rice-pile
experiments,
and also to Per Bak, the ultimate father of all this.
A.C.'s contribution was first financed by the TMR network on {\it Fractal
Structures and Self-Organization} through the EU-grant FMRX-CT98-0183.
Nowadays, he is happy to thank Santiago Ram\'on y Cajal
program to help Spanish young scientists.




\begin{references}
\bibitem{driving}
It is not important to keep a fixed tracer rate as long as
this quantity is much smaller than the main driving rate,
which is what sets the time scale.
For the same reason the tracer rate can vary for
different system sizes; 
in our case we have 1 or 2 min$^{-1}$,
which for the experimental resolution
is irrelevant in front of 43 min$^{-1}$.
\end{references}
\end{document}